\documentclass[aoas,preprint]{imsart}

\RequirePackage[OT1]{fontenc}
\RequirePackage{amsthm,amsmath}
\RequirePackage{natbib}
\RequirePackage[colorlinks,citecolor=blue,urlcolor=blue]{hyperref}
\usepackage{amssymb}
\usepackage{amsmath,epsfig, epsf, epic, graphicx} 
\usepackage{fancybox, lscape, subfigure} 
\usepackage{bm}
\usepackage{float}

\arxiv{arXiv:0000.0000}

\startlocaldefs
\numberwithin{equation}{section}
\theoremstyle{plain}

\endlocaldefs

\begin{document}

\begin{frontmatter}
\title{Spatial Multiresolution Analysis of the Effect of $\text{PM}_{2.5}$ on Birth Weights}
\runtitle{Spatial Multiresolution Analysis of $\text{PM}_{2.5}$ Grids}

\begin{aug}
\author{\fnms{Joseph} \snm{Antonelli}\thanksref{t1,m1}\ead[label=e1]{jantonelli@fas.harvard.edu}},
\author{\fnms{Joel} \snm{Schwartz}\thanksref{m1}\ead[label=e2]{joel@hsph.harvard.edu}}
\author{\fnms{Itai} \snm{Kloog}\thanksref{m2}\ead[label=e3]{ikloog@bgu.ac.il}}
\and
\author{\fnms{Brent} A. \snm{Coull}\thanksref{m1}
\ead[label=e4]{bcoull@hsph.harvard.edu}}

\thankstext{t1}{To whom correspondence should be addressed}
\runauthor{J. Antonelli et al.}

\affiliation{Harvard Chan School of Public Health\thanksmark{m1} and Ben-Gurion University of The Negev\thanksmark{m2}}

\address{Joseph Antonelli \\
Joel Schwartz \\
Brent A .Coull \\
Harvard T.H. Chan School of Public Health \\
655 Huntington Avenue\\
Boston, MA, 02115, USA \\
\printead{e1}\\
\phantom{E-mail:\ }\printead*{e2} \\
\phantom{E-mail:\ }\printead*{e4}}

\address{Itai Kloog\\
Department of Geography and Environmental Development \\
Ben-Gurion University of the Negev \\ 
Beer Sheva, Israel \\
\printead{e3}}
\end{aug}

\begin{abstract}
Fine particulate matter ($\text{PM}_{2.5}$) measured at a given location is a mix of pollution generated locally and pollution traveling long distances in the atmosphere. Therefore, the identification of spatial scales associated with health effects can inform on pollution sources 
responsible for these effects, resulting in more targeted regulatory policy.  Recently, prediction methods that yield high-resolution spatial estimates of $\text{PM}_{2.5}$ exposures allow one to evaluate such scale-specific associations. We propose a two-dimensional wavelet decomposition that alleviates restrictive assumptions required for standard wavelet decompositions. Using this method we decompose daily surfaces of $\text{PM}_{2.5}$ to identify which scales of pollution are most associated with adverse health outcomes. 
A key feature of the approach is that it can remove the purely temporal component of variability in $\text{PM}_{2.5}$ levels and calculate effect estimates derived solely from spatial contrasts.  This eliminates the potential for unmeasured confounding of the exposure - outcome associations by temporal factors, such as season.  We apply our method to a study of birth weights in Massachusetts, U.S.A from 2003-2008 and find that both local and urban sources of pollution are strongly negatively associated with birth weight. Results also suggest that 
failure to eliminate temporal confounding in previous analyses attenuated the overall effect estimate towards zero, with the effect estimate growing in magnitude once this source 
of variability is removed.
\end{abstract}

\begin{keyword}
\kwd{Wavelets}
\kwd{Multiresolution analysis}
\kwd{Spatiotemporal modeling}
\kwd{Environmental modeling}
\end{keyword}

\end{frontmatter}

\section{Introduction}

The epidemiologic literature investigating the health effects of air pollution is large, as countless studies have found associations between ambient levels of air pollution and a variety of adverse health outcomes (\cite{dockery1993association, samet2000fine, dominici2006fine}, with reviews of the literature provided by \cite{dominici2003health, pope2007mortality, breysse2013us}). Despite the large number of studies 
investigating the relationship between air pollution exposures and human health, there still exist critical unanswered questions that need to be addressed for the establishment of new regulations. Currently the U.S Environmental Protection Agency (U.S. EPA) regulates total $\text{PM}_{2.5}$ levels. However, $\text{PM}_{2.5}$ is generated by many different pollution sources. An important question is the extent to which various sources of pollution adversely affect health, knowledge of which could lead to targeted regulations more protective of human health. Establishing the extent to which locally generated pollution (such as that generated by traffic), urban background pollution, or long range transported pollution (such as that generated by coal fired power plants) is associated with health effects would be very useful in refining   air pollution standards in the future.   In this article we use the fact that locally generated traffic varies at fine spatial scales, urban background pollution induces city-to-city variability in pollution levels, and regional pollution that has travelled long distances varies very slowly over a region. 

There have been very few attempts to jointly model the health effects of regional, urban background, and local pollution sources, though it remains a crucial question in air pollution epidemiology. Black carbon is known to be highly associated with local traffic pollution. Sulfates are known to be spatially homogenous and represent pollution generated by coal fired power plants. 
\cite{maynard2007mortality} used a spatio-temporal prediction model  to predict black carbon levels for individuals in the greater Boston U.S.A area, and estimated the independent effects 
of black carbon and sulfates on mortality in a joint model. Other articles have decomposed air pollution levels into that generated by  local and regional sources without subsequently examining their respective effects on health. \cite{moreno2009profiling} examined the differences in hourly fluctuations of traffic and urban background components of PM10 in Santander, Spain. \cite{brochu2011development} used quantile regression to estimate the regional and local source contributions to black carbon levels in Boston and investigated how the sources of this pollutant  changed both across the year and within a given day. 

Due to recent advancements in $\text{PM}_{2.5}$ exposure estimation,  more complete data on  $\text{PM}_{2.5}$ levels beyond that provided by individual monitors are available, as remote sensing satellite data can now yield reliable $\text{PM}_{2.5}$ estimates  on a 1 $\times$ 1 km grid \citep{kloog2014new}. These new estimates of $\text{PM}_{2.5}$ are on a scale fine enough to allow novel approaches to spatial decomposition based on image analysis techniques. We show how such  decompositions of  the variability in  $\text{PM}_{2.5}$  levels across a region can yield insights into the sources of pollution most associated with health effect estimates. 
A variety of methods have been proposed to decompose images into different spatial scales, two of the most common techniques being wavelet decompositions and Fourier decompositions. For the remainder of the manuscript we focus on wavelets.   Due to the existence of point and line sources of pollution, such as interstates and other roadways, the surface of $\text{PM}_{2.5}$ will contain many spikes. Wavelets are well known to be a useful basis function for preserving sharp features in signals \citep{petrosian2013wavelets}, and many spike detection algorithms are based on wavelet transforms \citep{hulata2002method, nenadic2005spike}. One of the main goals of the study is to characterize the impact of traffic pollution on health and therefore it is important to adequately capture, and not oversmooth, these spikes in pollution. Moreover, wavelets decompose a spatial surface into multiple spatial scales that are orthogonal, which allows us to avoid multicollinearity among multiple scales and the resulting instability of effect estimates from a health effects model.

In this paper we focus on estimating associations between different spatial scales of $\text{PM}_{2.5}$ and birth weight in Massachusetts, U.S.A.  A challenge in quantifying the health effects of air pollution on birth outcomes is the potential for temporal confounding due to the fact that births are not distributed uniformly over the course of the year, and there can be seasonal variability in both pollution levels and  birth outcomes \citep{darrow2009seasonality}.
One approach to accounting for factors that vary temporally and affect both exposure and outcome is to control for seasonality in the model.  However, for this approach to remove
bias arising from temporal confounding completely, one needs the model to be correctly specified and this can be difficult to achieve in some settings \citep{peng2006model}.  Another approach is to condition out, or remove, the temporal component of exposure variability. 
Another feature of our proposed approach is that by applying the proposed spatial decompositions on a day-by-day basis, each of which contains a separate term representing the overall mean $\text{PM}_{2.5}$ on a given day, we partition the variability in $\text{PM}_{2.5}$ into purely temporal and multiple spatial components.  This partitioning eliminates the possibility that the health effect estimates associated with the spatial contrasts at different spatial scales are affected by temporal confounding, which can be very difficult to 
completely eliminate through modeling.  Our analyses of the association between $\text{PM}_{2.5}$ exposures and birth weights in Massachusetts from 2003-2008 show that 
failure to fully eliminate the effects of temporal confounders in previous analyses may have attenuated the overall effect estimate towards zero, with the effect estimate growing in magnitude once this source 
of variability is removed.

Standard two-dimensional wavelet decompositions typically require that the surface being decomposed is rectangular and the points are uniformly spaced. An additional requirement of standard wavelet analysis is that the points are dyadic, meaning that the number of points on the surface grid is $2^l$, where $l$ is some positive integer, although ``padding", the practice of adding points to the surface, can be used to satisfy this requirement. Our interest focuses on decomposing a spatial surface of $\text{PM}_{2.5}$ across New England, U.S.A., a setting in which none of these conditions are met. There is no reason to think our data would be dyadic, and the satellite data yielding the $\text{PM}_{2.5}$ estimates are not laid out on a perfectly uniform grid. Previous work has avoided both the dyadic and uniform grid assumptions through a variety of techniques, such as the lifting scheme and interpolation \citep{sweldens1998lifting, xiong2006lifting, pollock2007non, gupta2010non}. Others have generalized wavelet theory using radial basis functions, which are not constrained to lie on a uniform grid \citep{buhmann1995multiquadric, chui1996wavelets}. To the best of our knowledge, however,  none of these have provided a practical way of decomposing a surface that is not rectangular. In this paper we develop  a two-dimensional extension to work on a penalized regression representation of wavelets, originally proposed in \cite{wand2011penalized}, which relaxes these assumptions  of a standard wavelet analysis. The advantages of this  method include its simple application, its ability to scale to large spatial surfaces, and its ability to avoid restrictive assumptions about the shape or features of the surface. 

In this paper, we use the proposed wavelet based method to decompose daily surfaces of $\text{PM}_{2.5}$ across New England and use the components of the resulting decomposed surface as covariates in a health effects model relating  birth weights in Massachusetts to scale-specific $\text{PM}_{2.5}$.  Section \ref{sec:birthsmotivating} introduces the pollution data and motivating scientific problem. Section \ref{sec:decomp} introduces the proposed method for performing 2d wavelet decompositions on irregular grids. Section \ref{sec:waveletapp} illustrates the decomposition in the $\text{PM}_{2.5}$ data. Section \ref{sec:birthsanalysis} applies the method to analyze the association between scale-specific $\text{PM}_{2.5}$ and birth weights in Massachusetts, and Section \ref{sec:waveletdiscussion} concludes with further discussion.

\section{$\text{PM}_{2.5}$ and Birth Weights in Massachusetts}
\label{sec:birthsmotivating}

\subsection{Exposure Data}

Typically $\text{PM}_{2.5}$ is measured at monitoring stations, which are located sporadically across the United States. In early health effect studies, conditional on the monitoring data, $\text{PM}_{2.5}$ exposure for an individual in a given location was assigned the value from the nearest monitor or a weighted average of monitors within a pre-defined range. In recent years monitoring data has been augmented with geographical and remote sensing information to yield  individual, residence specific estimates of $\text{PM}_{2.5}$ levels.  Specifically, in previous work we have combined ideas from land use regression and mixed models, and incorporated satellite aerosol optical depth (AOD) measurements to obtain widespread estimates of $\text{PM}_{2.5}$ at a 1 $\times$ 1 km resolution \citep{kloog2014new}. Satellite AOD is a measure of light attenuation in the atmospheric column that is affected by ambient conditions and can be used to help estimate $\text{PM}_{2.5}$. Satellite estimates of $\text{PM}_{2.5}$ on a 1 $\times$ 1 km grid are available daily from 2003 to 2011 for the Northeastern United States and they give an accurate estimate of the surface of $\text{PM}_{2.5}$ in this area, as judged by cross-validation. Specifically, \cite{kloog2014new} showed the $R^2$ values between predictions and true values observed at monitors is around 0.9, indicating high predictive accuracy in locations in which monitoring data exist. The right panel of Figure \ref{fig:3plot} presents air pollution exposure estimates for the study region of interest.
The 1 $\times$ 1 km scale of the estimated exposures allows us to apply our proposed decomposition method and examine the effects of air pollution across a range of spatial scales.

\subsection{Birth Weights}

Many epidemiological studies have established relationships between $\text{PM}_{2.5}$ and adverse birth outcomes. \cite{glinianaia2004particulate} and \cite{dadvand2013maternal} provided a review of the literature. In Massachusetts, \cite{kloog2012using} reported an association between $\text{PM}_{2.5}$ and birth weights  using Satellite AOD based $\text{PM}_{2.5}$ estimates on a 10 $\times$  10 km grid. We extend this work by using the finer scale, 1 $\times$  1 km satellite based $\text{PM}_{2.5}$ estimates as well as estimating associations between birth weight and specific spatial scales of variation of $\text{PM}_{2.5}$ exposure. \cite{kloog2012using} provided specific details on the birth weight data.  Briefly, the study population includes all singleton live births from the Massachusetts Birth Registry from January 1st, 2000 to December 31st, 2008. We restrict attention to births after October 1st, 2003 as the satellite based $\text{PM}_{2.5}$ data is only available from 2003 onwards. The data set contains 332,717 singleton births with the geocoded address of each mother at the time of birth and potential confounders.

\section{Wavelet Decomposition for Irregular Grids}
\label{sec:decomp}
\subsection{1d Wavelet Analysis on Irregular Grids}

To motivate our approach and establish notation, we start by reviewing the penalized regression representation of a wavelet decomposition proposed by \cite{wand2011penalized} for a one-dimensional functional response. This formulation has the advantage over a standard wavelet decomposition that it removes the requirement that the data lie on a uniform grid. We will then extend this penalized regression approach to the two-dimensional setting. In the one-dimensional case, suppose we have data, $y_i$ for $i=1,\dots,n$ observed at locations $x_i$, with no restrictions on the spacing or the dimension of $\boldsymbol{x}$. We represent the data $\boldsymbol{y}$ as 

\begin{align}
	y_i &= f(x_i, \boldsymbol{\theta}) = \theta_0 + \sum_{k=1}^{K} \theta_k z_k(x_i),
\end{align}

\noindent where $z_k(\cdot)$ are wavelet basis functions and $\theta_k$ are wavelet coefficients. The wavelet coefficients have a nice interpretation in terms of scale and location of the signal. At level $l$ of a wavelet decomposition there are $2^l - 1$ basis functions. The lower level basis functions capture low frequency signals, while the higher level basis functions capture the higher frequency signals, in the data. In terms of the motivating $\text{PM}_{2.5}$ application,  the lower level functions will capture smooth, regionally varying trends in pollution, while the higher level functions will capture fine scale  $\text{PM}_{2.5}$ spatial variation that captures differences in locally generated pollution such as that generated by traffic.  In this representation,  $z_1(\cdot)$ is the basis function for the first wavelet level, $z_2(\cdot) \ \text{and} \ z_3(\cdot)$ are the basis functions at the second wavelet level, and so on for levels $3$ to $L$.

Classical wavelet transforms are defined on equally-spaced grids on the unit interval, $[0,1)$. When the data are dyadic and regularly spaced on the unit interval, the wavelet basis functions, which we
denote $z_k^u(\cdot)$,  are tractable and the discrete wavelet transform, a fast $O(n)$ algorithm, can be used to estimate $\widehat{\boldsymbol{\theta}}$. We are, however, in the setting of unequally spaced data, in which wavelet basis functions are not tractable. To solve this issue we first define wavelet functions as

%\begin{align}
%\left[1 \ z_1^u(\frac{i-1}{R}) \ \dots \ z_{R-1}^u(\frac{i-1}{R})\right].
%\end{align}
%
%\noindent In this case,  the orthogonality of W yields estimates of $\theta$ 
%
%\begin{align}
%	\widehat{\theta} = W^T y.
%\end{align}
%
%\noindent The Discrete Wavelet Transform, a fast $O(R)$ algorithm, can be used to estimate $\widehat{\theta}$ in this setting. 
%
%\subsection{Extension to irregular grids}
%
%We will now review extension of the above formulation to the case where the data is not dyadic or on a regular grid. Now consider data, $y_i$ for $i=1,\dots,n$ observed at locations $x_i$, with no %restrictions on the spacing or the dimension of $y$ {\bf CITE WAND PAPER AGAIN HERE}. Define a new set of basis functions as
%
\begin{align}
z_k(x_i) = z_k^u\left(\frac{x_i-a}{b-a}\right), \ \ \ k = 1,\dots,K,
\end{align}

\noindent  where $a$ and $b$ are the minimum and maximum of $\boldsymbol{x}$ respectively.  We then estimate $z_k^u(\cdot)$ at arbitrary locations on the unit interval, since our data does not fall on an equally spaced grid. To do this, \cite{wand2011penalized} defined a very fine grid of points on the unit interval, with the number of points in our grid being a multiple of 2, and evaluated the basis functions at each point on this grid.  This grid needs to include a large number of points so that any of the observed data points will lie very closely between two grid points. One can then calculate the value of $z_k^u(\cdot)$ at any point in the interval as a linear interpolation of the two nearest grid points. 
%
%That is, 
%
%\begin{align}
%z_k^u(x) \approx \left\{1 - (xR - \lfloor xR \rfloor) \right\} z_k^u \left(\frac{\lfloor xR \rfloor}{R} \right) + (xR - \lfloor xR \rfloor) z_k^u \left(\frac{\lfloor xR \rfloor + 1}{R} \right).
%\end{align}

\noindent  A variety of standard wavelet basis functions can be used to form $z_k^u (\cdot)$ \citep{torrence1998practical}, and for this paper we use the Debauchies 5 wavelet \citep{daubechies1988orthonormal}. Upon defining wavelet basis functions at any given $\boldsymbol{x}$ value, one can estimate the wavelet coefficients $\boldsymbol{\theta}$. In this work we will use the LASSO, which  estimates the coefficients using an L1 penalty on $\boldsymbol{\theta}$,  to select only non-zero coefficients.  The solution satisfies 

\begin{align}
	\widehat{\boldsymbol{\theta}} = \underset{\boldsymbol{\theta}}{\text{arg min}} \left\{\sum_{i=1}^n \left(y_i - f(x_i, \boldsymbol{\theta}) \right)^2 \right\} \text{ subject to } \sum_j |\theta_j| < c,
\end{align}

\noindent where $c$ is a tuning parameter that controls the amount of penalization \citep{tibshirani1996regression}.  Variation in $\boldsymbol{y}$ at a given level then corresponds to the product of the wavelet coefficients and the basis functions from that particular level.   Because our interest focuses on multi-resolution analysis of a two-dimensional image of $\text{PM}_{2.5}$, we consider a two-dimensional extension of this approach, which we outline in the next section. 

\subsection{Two-dimensional Wavelets on an irregular grid}

Suppose now that we have $n$ data points, $y_i^*, \ i=1,\dots,n$ that lie on a two-dimensional space indexed by location $(x_{1i}, x_{2i})$. In the air pollution application, $\boldsymbol{y^*}$ is the $\text{PM}_{2.5}$ level at (longitude, latitude) location $(\boldsymbol{x_1}, \boldsymbol{x_2})$ on any given day.   We first apply the wavelet approach for one dimension, as outlined in  the previous section, to each dimension separately.  That is, 
we define basis functions 

\begin{align}
z_k^{x_1}(x_{1i}) = z_k^u\left(\frac{x_{1i}-a_1}{b_1-a_1}\right), \ \ \ k = 1,\dots,K, \\
z_k^{x_2}(x_{2i}) = z_k^u\left(\frac{x_{2i}-a_2}{b_2-a_2}\right), \ \ \ k = 1,\dots,K,
\end{align}

\noindent where $a_1$ and $b_1$ define the range of $\boldsymbol{x_1}$, and $a_2$ and $b_2$ the range of $\boldsymbol{x_2}$. We construct two-dimensional wavelet basis functions as tensor products of the one-dimensional functions in each direction. The two-dimensional basis functions can be written as

\begin{align}
	z_{l,m}(x_{1i}, x_{2i}) = z_l^{x_1}(x_{1i}) z_m^{x_2}(x_{2i}).
\end{align}

\noindent The model now takes the form 

\begin{align}
	y_i^* &= f(x_{1i}, x_{2i}, \boldsymbol{\theta}) = \theta_0 + \sum_{l=1}^{K} \sum_{m=1}^K \theta_{lm}z_{l,m}(x_{1i}, x_{2i}),
\end{align}

\noindent where we again estimate $\boldsymbol{\theta}$ using the LASSO. The 2d wavelet basis functions are products of 1d basis functions from each direction, which can be of a different spatial scale. 
Therefore, 2d basis functions formed as the product of 1d basis functions from the same scale have square support, whereas 2d basis functions formed as the product of 
two basis functions from different spatial scales have rectangular support.  Therefore,  this construction creates some basis functions with elongated support, which differs from the commonly used square wavelet transform. We do not view this as a crucial distinction in the context of decomposing $\text{PM}_{2.5}$ in Massachusetts, particularly because we will be looking at exposures such as the high frequency component that is the sum over multiple wavelet levels. If our interest were in individual wavelet levels, then this difference would be of more importance.

\section{Wavelet-based Decomposition of New England $\text{PM}_{2.5}$ Data}
\label{sec:waveletapp}

We now  apply the proposed method for two-dimensional wavelet decomposition on irregular grids to $\text{PM}_{2.5}$ data in New England, U.S.A.,  separately for each day. Previous work in air pollution exposure assessment has established that spatial variation in $\text{PM}_{2.5}$ levels can be decomposed into three primary spatial scales: large scale regional variation, city-to-city variation (urban background), and localized variation (such as that generated by traffic). Within the study region of interest, $\text{PM}_{2.5}$ levels attributable to regional sources vary little within a given day, but may vary greatly from day to day. This suggests that changes in pollution generated by regional sources will be captured by temporal variation in $\text{PM}_{2.5}$, while changes in pollution generated by urban and local traffic sources will be captured by spatial variability. That is, variability generated by urban pollution sources will be captured by lower level wavelet coefficients as they vary over longer distances, whereas variability generated by local pollution sources will be captured by the higher level wavelet coefficients. 

The proposed method scales quite well to large surfaces. We applied the method to $\text{PM}_{2.5}$ surfaces that contained approximately 70,000 grid points. We use seven wavelet levels in each direction, which yields $2^{14} = 16,348$ basis functions and a regression model with a design matrix of dimension 70,000 $\times$ 16,384.  We apply the LASSO to fit the model  using the \texttt{glmnet} package in R \citep{glmnet}, applying cross validation to select the LASSO tuning parameter for each day separately. 

\begin{figure}[h]
\centering
 \includegraphics[width=0.95\linewidth]{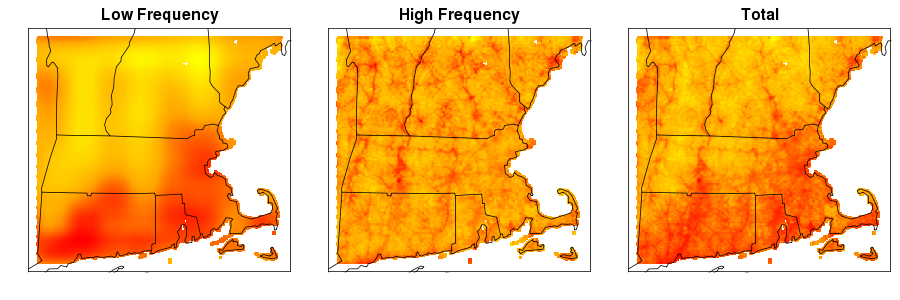}
\caption{Illustration of average wavelet decomposition of the satellite-derived $\text{PM}_{2.5}$ estimates,  averaged over all days in 2006. The left panel shows the low frequency component from a wavelet decomposition, the middle panel is the high frequency component from a wavelet decomposition, and the right panel is the total surface.}
\label{fig:3plot}
\end{figure}

One issue with defining low and high frequency scales of spatial variation from the fit of the model with seven wavelet levels is 
the selection of a cutoff that classifies a wavelet level as low frequency or high frequency.   In Figure \ref{fig:3plot} we defined low frequency variation to include basis functions from wavelet levels 1 through 3 in both longitude or latitude. We selected this threshold by visual inspection of the decomposed surfaces, which suggested that the resulting decomposition best represented smooth variation that best corresponded with existing knowledge of spatial patterns and pollution sources in the New England region.  This choice defines the high frequency component of variation, representing pollution generated by local sources,  as containing wavelet levels 4 through 7 in either the longitude or latitude direction. 

Figure \ref{fig:3plot} illustrates the average of the low frequency and high frequency components of $\text{PM}_{2.5}$ across each day in 2006 for New England. The figure shows how the pollution surface can be decomposed into two separate components, representing different spatial frequencies at which pollution varies. To calculate the low frequency component, we create a new vector of coefficients 
$\tilde{\boldsymbol{\theta}}$ which is equal to $\widehat{\boldsymbol{\theta}}$ upon setting the coefficients corresponding to higher frequency basis functions equal to zero, and calculate $f(\boldsymbol{x_1, x_2, \tilde{\theta}})$. To calculate the  high frequency component, we take the difference between the true surface and the lower frequency component as the higher frequency component.

\section{Analysis of Scale-Specific Associations Between $\text{PM}_{2.5}$ and Birth Weight}
\label{sec:birthsanalysis}

We apply the proposed decomposition to examine the association between variability in $\text{PM}_{2.5}$ levels  at different spatial scales and birth weights in Massachusetts, U.S.A. for the period 2003-2008. We perform the wavelet decomposition for each day during the study period to obtain pollution surfaces representing different spatial scales. All confidence intervals have been adjusted for multiple comparisons using a Bonferroni correction within each model. 

\subsection{Low Versus High Frequency Components}
\label{sec:lowVShigh}

We first examine the independent associations between each of low and high frequency variation in $\text{PM}_{2.5}$ and birth weight, using the definitions of low and high frequency variation provided in the previous section.   When birth weight is the outcome, there are several potentially relevant exposure windows:  the full gestation period, a given trimester, or the last 30 days of the gestation period. In this work, we report results on associations between birth weights and the trimester-specific exposures.  Because births are not distributed uniformly within a year and because birth outcomes can also vary within a year, there also exists a purely temporal component of $\text{PM}_{2.5}$ variability that could potentially be associated with birth weight. Due to this fact, we decompose our exposure surface for each day into three separate components: A component that is simply the mean $\text{PM}_{2.5}$ across the region on any given day, a low frequency spatial component, and a high frequency spatial component. For each  infant in the study, we compute each of these three exposure components  for each day based on the residential location of the mother as denoted on the birth certificate, and then average these contributions across the time period corresponding to a particular trimester of interest.  Summation of the daily means across the exposure window of interest captures purely temporal variation in $\text{PM}_{2.5}$ levels, due to the fact that women have babies on different days, whereas the low and high frequency spatial scales capture variability in exposure due to the fact that mothers live in different spatial locations. 

We examine the effects of $\text{PM}_{2.5}$ at different scales using  two models. The first model is

\begin{align}
	BW_{ij} = (\beta_0 + \beta_{0j}) + \beta_1 {\text{PM}_{2.5}}_{ij} + \boldsymbol{\beta}_c\textbf{C}_{ij} + \epsilon_{ij},
	\label{eqn:modKloog}
\end{align}

\noindent where $BW_{ij}$ represents the observed birth weight,  $ {\text{PM}_{2.5}}_{ij}$ is the estimated total $\text{PM}_{2.5}$ value for a given trimester, and 
 $\textbf{C}_{ij}$ contains a set of potential confounders, for subject $i$ in census tract $j$. These confounders included gestational age, mothers age, cigarette use, income, education, chronic conditions of the mother, pregnancy conditions, previous occurrence of a preterm birth, and percent open space of a census tract. We include a random intercept for census tract to control for correlation among infants  in the same census tract. 
 The residual errors, $\epsilon_{ij}$, are independent, mean zero normal random variables.  \cite{kloog2012using} provided justification for this model and the confounders included in it.  
 This previous work does not use  the spatial decompositions of the $\text{PM}_{2.5}$ exposures. Therefore,  $\beta_1$ represents the  effect of total $\text{PM}_{2.5}$ exposure, which blends associations between the temporal and multiple spatial scales of exposure.  

\begin{figure}[h]
\centering
	\includegraphics[height=4in]{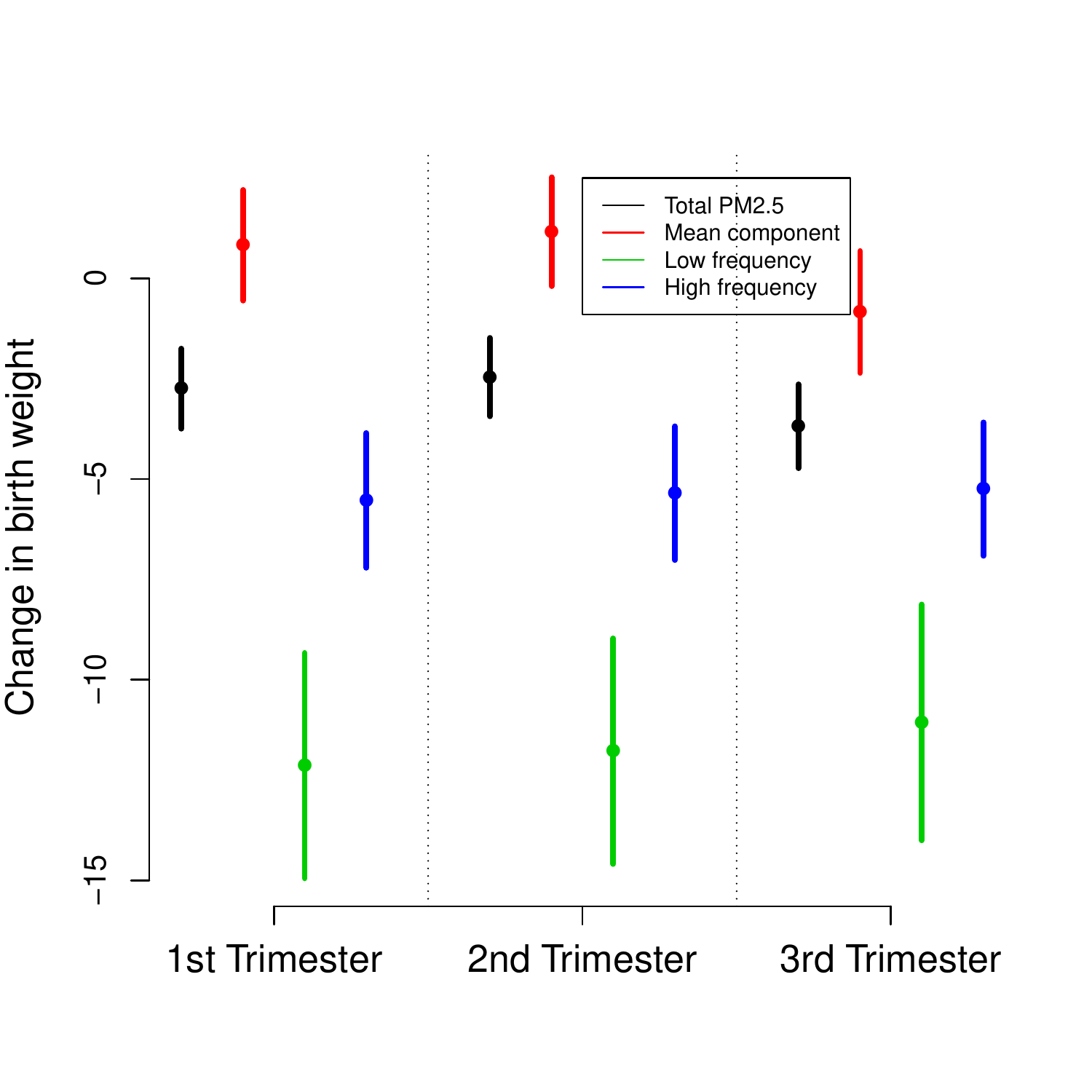}
\caption{Parameter estimates and corresponding 95\% confidence intervals from $\text{PM}_{2.5}$ models for each time period. The black lines represent the point estimates and 95\% confidence intervals for 
$\beta_1$ from model \ref{eqn:modKloog} and the remaining colors represent those for  $\beta_1^*,  \beta_2^*,  \beta_3^*$ from model \ref{eqn:modJoey}}
\label{fig:modelEst}
\end{figure}

To examine how different spatial scales and temporal variation in $\text{PM}_{2.5}$ exposure are associated with birth weight we fit the model 

\begin{align}
	BW_{ij} = (\beta_0^* + \beta_{0j}^*) + \beta_1^* \text{Mean}_{ij} + \beta_2^* \text{Low}_{ij}  + \beta_3^* \text{High}_{ij} + \boldsymbol{\beta}_c^* \textbf{C}_{ij} + \epsilon_{ij},
	\label{eqn:modJoey}
\end{align}

\noindent where $\text{Mean}_{ij}$,   $\text{Low}_{ij}$, and  $ \text{High}_{ij}$ represent the temporal, low spatial frequency, and high spatial frequency components of $\text{PM}_{2.5}$ variability, respectively.  The confounders, random intercepts, and residuals are defined as above.  
Figure \ref{fig:modelEst} shows the results from both model (\ref{eqn:modKloog}) and (\ref{eqn:modJoey}) for  each of the trimesters. 

The results indicate that the magnitude and direction of the effect estimates, and therefore the patterns in effect estimates across components of  $\text{PM}_{2.5}$ variation, are similar across trimesters.  Both the low and high frequency components of $\text{PM}_{2.5}$ are strongly negatively associated with birth weight. Interestingly the mean component exhibits very small, and in the case of trimesters 1 and 2 even slightly positive, associations with birth weight.  One potential explanation for this result is that the mean component represents variation in  $\text{PM}_{2.5}$ that is subject to temporal confounding. Both $\text{PM}_{2.5}$ and birth weights decrease over time during the study period, which could explain the  observed positive association. To test this hypothesis we fit the model  \ref{eqn:modJoey} using first trimester exposure, but included a smooth term for date into the vector of potential confounders, $\textbf{C}_{ij}$. After applying this model the effect of the mean component decreases to -2.7 g (-6.4 g, 1.1 g), indicating that the original estimate is confounded by temporal factors.  By removing this variability in  $\text{PM}_{2.5}$  from the low and high frequency spatial scales, we eliminate the possibility of purely temporal factors confounding the effect estimates for the spatial contrasts in exposure.  The resulting effect estimates for the low and high frequency spatial component $\text{PM}_{2.5}$ are both larger than the overall $\text{PM}_{2.5}$  effect estimate, $\widehat{\beta}_1$, and this is likely because we have removed the temporal sources of variation that bias the overall estimate towards the null. This analysis also provides strong evidence that the effect estimate for the low frequency component is larger  than that for the high frequency component (p-value $< 0.0001$).

\subsection{Removal of Fine Scale Variation}

As a secondary analysis, we now investigate which wavelet levels of $\text{PM}_{2.5}$ are driving the association between total $\text{PM}_{2.5}$ and birth weight. To this end,  we repeatedly fit the model with a
single pollution exposure in the model, after successively removing high frequency wavelet levels from the daily pollution surfaces. We also keep the mean component of $\text{PM}_{2.5}$ variability separate from the spatial component,   as the results from the previous section indicate that it could attenuate the overall association between $\text{PM}_{2.5}$ and birth weight. For a given trimester, the model of interest is now

\begin{align}
	BW_{ij} = (\tilde{\beta_0} + \tilde{\beta_{0j}}) + \tilde{\beta_1} \text{Mean}_{ij} + \tilde{\beta_2} {\text{PM}_{2.5}^R}_{ij} + \boldsymbol{\tilde{\beta}}_c \textbf{C}_{ij} + \epsilon_{ij}, 
	\label{eqn:modScales}
\end{align}

\noindent where ${\text{PM}_{2.5}^R}_{ij}$ is the spatial component of  $\text{PM}_{2.5}$ variability after removing a certain number of higher wavelet levels from that component.   

\begin{figure}[h]
\centering
	\includegraphics[width=0.95\linewidth]{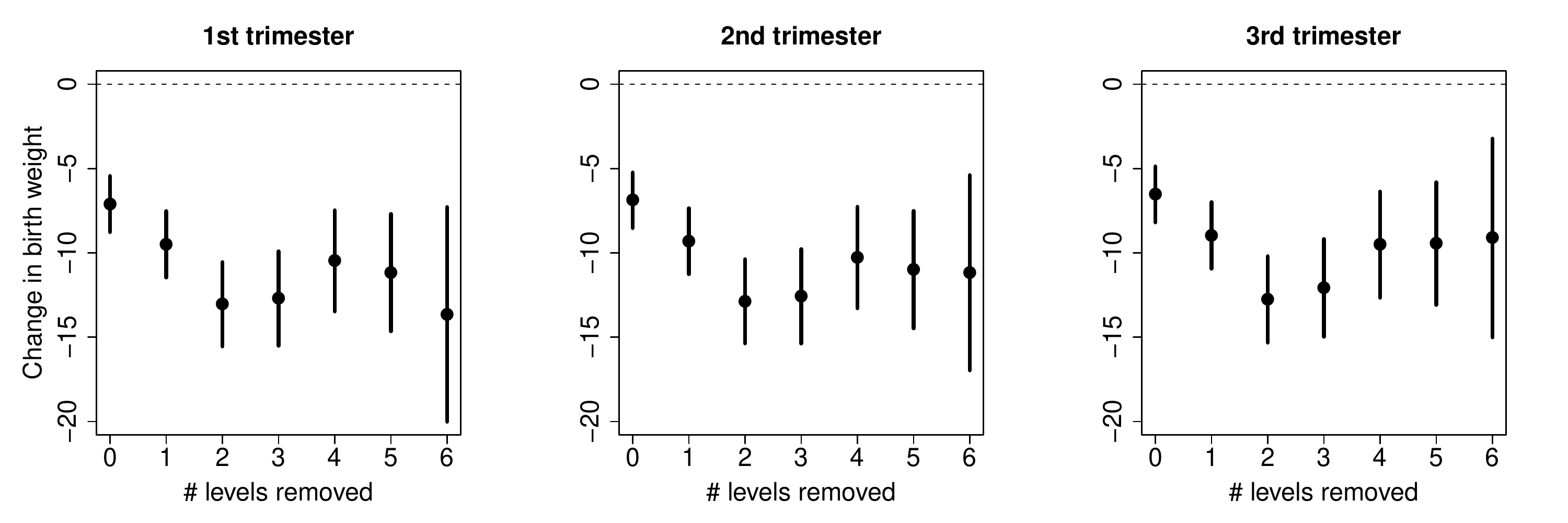}
\caption{Parameter estimates and corresponding 95 \% confidence intervals from model \ref{eqn:modScales} when we remove high frequency spatial levels for each trimester. Within each panel from left to right we successively remove more and more of the higher frequency levels}
\label{fig:removeScales}
\end{figure}

This models differs from  \ref{eqn:modKloog} in two ways. First we include the mean component into the model to partition the potential sources of temporal confounding from the spatial contrast. 
  We assess how  the estimate  of $\tilde{\beta}_2$ changes as 
   the high frequency wavelet levels are removed one at at time. Figure \ref{fig:removeScales} shows the estimates and 95\% confidence intervals from model \ref{eqn:modScales} as we successively remove high frequency information. 

Results from the different trimester-specific analyses show a similar  pattern. Upon removal of the highest 1st and 2nd wavelet levels from the $\text{PM}_{2.5}$ surface, 
  the magnitude of the  $\text{PM}_{2.5}$ effect estimate  increases, with the effect estimate changing from -6.5 g (-8.2 g, -4.9 g) to -12.7 g (-15.3 g, -10.2 g) for the 3rd trimester. This result suggests that the effects at these very high frequency levels are smaller in magnitude than their lower frequency counterparts. These levels, which are largely spatially uncorrelated variation in exposure, likely contain exposure measurement error. Therefore one would expect that filtering them out would yield stronger estimated associations between exposure and outcome.  As additional levels are removed from the total $\text{PM}_{2.5}$ exposure, there is a decline in effect size, with a visible change occurring at wavelet level 4. For the 3rd trimester analysis,  the effect estimate drops in magnitude from -12.1 g (-15.0 g, -9.2 g) to -9.5 g (-12.7 g, -6.4 g) when we exclude the fourth level from total $\text{PM}_{2.5}$.  The effect estimates also change noticeably when one includes the 1st level into the model. The $\text{PM}_{2.5}$ effect estimate when $\text{PM}_{2.5}$ contains only the first wavelet level is -9.1 g (-15.0 g, -3.2 g) for trimester 3, and  -13.6 g (-20.0 g, -7.3 g) for trimester 1, suggesting a strong association between outcome and $\text{PM}_{2.5}$ at this level. Therefore, the first level appears to drive the health effects observed in the analyses presented in Section \ref{sec:lowVShigh}, whereas the fourth level appears to drive the health effect associated with the high frequency component in that analyses. Further,  spatially uncorrelated measurement error contained in the sixth and seventh wavelet levels of variability appear to attenuate the high frequency effect estimate reported in  Section \ref{sec:lowVShigh}. 

\section{Discussion}
\label{sec:waveletdiscussion}

In this article we decomposed daily 1 $\times$ 1 km estimated concentrations of  $\text{PM}_{2.5}$  and then examined how different spatial  scales are associated with birth weights in Massachusetts U.S.A.  A key feature of the approach is that, because we applied the spatial decomposition on a day-by-day basis, we removed the purely temporal component of variability in $\text{PM}_{2.5}$ levels and calculated effect estimates derived  from spatial contrasts.  This approach eliminates the potential for unmeasured confounding of the exposure - outcome associations by temporal factors, such as seasonality. The resulting estimate of the association between the outcome and the temporal component of $\text{PM}_{2.5}$ was positive or close to zero for each trimester window of exposure, suggesting that the overall estimate of an $\text{PM}_{2.5}$ effect is attenuated by temporal confounding. Further, we estimated that variation in pollution at both low and high spatial scales was significantly negatively associated with birth weight in Massachusetts, with the  low frequency association larger than the high frequency association. We also examined the effect of removing variation in $\text{PM}_{2.5}$ captured by individual wavelet levels one at a time and estimated that removal of very high frequency variation, which was represented by the sixth and seventh levels, actually increased the association between $\text{PM}_{2.5}$ exposure and birth weight.  This could suggest that this very fine scale, spatially uncorrelated variability in exposure is measurement error associated with the AOD 
measurements used as inputs in the exposure prediction algorithm.  Wavelets have been used to de-noise signals and removal of the sixth and seventh scales can be thought of as a form of de-noising, thereby potentially removing the exposure error and increasing the magnitude of the effect estimates associated with the remaining variation in exposure. 

We have proposed a two-dimensional wavelet decomposition that is flexible, easy to implement, and scalable to large spatial surfaces. By extending ideas from \cite{wand2011penalized} to spatial settings,  we have employed an image decomposition that does not rely on assumptions of standard wavelet theory that are overly restrictive. This approach places a wavelet decomposition within a penalized regression framework that simplifies implementation and estimation of wavelet coefficients. The proposed method will allow researchers to perform multi-resolution analyses of spatial data regardless of the structure and scale of their data. 

One limitation of the results from the study of birth weights is that we are ignoring the fact that the $\text{PM}_{2.5}$ measurements are estimated, leading to some uncertainty that is not incorporated into our analyses. The confidence intervals placed on our model estimates are under the assumption that the estimated $\text{PM}_{2.5}$ concentrations are known, which could lead to interval estimates that are anti-conservative. \citep{alexeeff2015consequences} characterized scenarios in which this uncertainty is likely to have a
meaningful impact on inference and when it is not.  These authors showed that, while inferences on health effect estimates of short-term (e.g. daily)
 exposures estimated by relatively simple spatial prediction models (e.g. kriging) can be severely 
anti-conservative, inference on effects of chronic health effects associated with long-term average
exposures generated by more complex, well-fitting land-use spatial regression models are less affected.  Specifically, 
in this latter case 95\% confidence interval coverages were only slightly lowered (in the range of 91 to 94\% depending on the number of monitors used). 
Since interest in the present study focuses on long-term exposures during pregnancy, or a trimester of pregnancy, estimated by a relatively 
complex hybrid spatio-temporal model based on a combination of land-use and remote-sensing satellite data, we interpret these previous findings 
to imply that these inferences will be only slightly anti-conservative and therefore not such a significant factor in the current study.

A related limitation is that because we are using estimates of $\text{PM}_{2.5}$ there might be measurement error that biases the effect estimates of interest. 
While we likely remove some of the measurement error when removing the highest two frequency wavelet scales, it's possible that measurement error induces bias in our effect estimates. Future work could focus on applying well developed measurement error correction techniques within the multi-resolution spatial analyses to examine the impact of this error. A final limitation is that our approach does not retain the computational speed of the discrete wavelet transform. The approach fits a penalized regression model of dimension $n$ by $K^2$, which can be quite large in practical applications. This is a price we pay to generalize the model to handle irregular grids.  However, we have not found this limitation to be a serious problem in the regionally-sized application such as this one,  as we were able to 
apply our approach to spatial surfaces of $n=70,000$ data points and $K^2 = 16,384$ basis coefficients. 

There are many possibilities for future research directions. One  could use the wavelet decompositions of $\text{PM}_{2.5}$ to learn more about variation in specific components of 
$\text{PM}_{2.5}$. Data is available at monitoring sites about these components, which means that we can explore correlations between the decompositions and $\text{PM}_{2.5}$
component data at monitoring locations, which could provide additional insights into the specific pollution sources that are captured at a given spatial scale.  It would also be of interest to examine whether other health endpoints representative of different biologic mechanisms are associated with specific scales of variation in $\text{PM}_{2.5}$ levels.

\section*{Acknowledgements}

This study was supported by grants from the National Institutes of Health  (ES000002, ES024332, ES007142) and  U.S. Environmental Protection Agency grant RD-834798-01.

\bibliographystyle{imsart-nameyear}
\bibliography{decomposition}

\end{document}